\documentclass[letter]{aa}
\usepackage[varg]{txfonts}
\usepackage{graphicx}
\usepackage{natbib}
\bibpunct{(}{)}{;}{a}{}{,} 

\begin{document}

\title{Solar surface rotation: N-S asymmetry and recent speed-up}

\titlerunning{Rotation of ALs of solar X-ray flares}

%
%

\author{L. Zhang \inst{1,}\inst{2} , K. Mursula\inst{1}, \and I. Usoskin\inst{1,}\inst{3}}

\institute{ReSoLVE Centre of Excellence, Department of Physics,
University of Oulu, Finland,  \email{liyun.zhang@oulu.fi}
\and Key Laboratory of Solar Activity, National Astronomical Observatories,
Chinese Academy of Sciences, Beijing, China
\and  Sodankyl{\"a} Geophysical Observatory, University of Oulu, Finland }

%

\abstract
{The relation between solar surface rotation and sunspot activity still remains open.
Sunspot activity has dramatically reduced in solar cycle 24 and several solar activity indices and flux measurements experienced unprecedentedly low levels during the last solar minimum.}
{We aim to reveal the momentary variation of solar surface rotation, especially during the recent years of reducing solar activity.}
{We used a dynamic, differentially rotating reference system to determine the best-fit annual values of the differential rotation parameters of active longitudes of solar X-ray flares and sunspots in 1977-2012.}
{The evolution of  rotation of solar active longitudes obtained with X-ray flares and with sunspots is very similar.
Both hemispheres speed up since the late 1990s, with the southern hemisphere rotating slightly faster than the north.
Earlier, in 1980s, rotation in the northern hemisphere was considerably faster, but experienced a major decrease in the early 1990s.
On the other hand, little change was found in the southern rotation during these decades.
This led to a positive asymmetry in north-south rotation rate in the early part of the time interval studied.}
{The rotation of both hemispheres has been speeding up at roughly the same rate since late 1990s, with the southern hemisphere rotating slightly faster than the north.
This period coincides with the start of dramatic weakening of solar activity, as observed in sunspots and several other solar, interplanetary and geomagnetic parameters.}


\maketitle


\section{Introduction}
\label{Introduction}

A number of solar activity indices and flux measurements reached abnormally low values during last solar minimum, which was also exceptionally long.
The activity level in the ongoing cycle 24 dramatically reduced compared to previous few solar cycles.
 The remarkably long and deep solar minimum and the weak cycle 24 have caused intensive attention in the solar and space physics community \citep{jianetal2011, cletteandlefevre2012, wangetal2009, solomonetal2010, solomonetal2011, cliverandling2011}.

 Here we study the changes in solar surface rotation during the last few decades, including the period of activity weakening, by analyzing the rotation of solar active longitudes.
Sunspots and other forms of solar magnetic activity are not uniformly distributed in solar longitude, but are centered around certain longitude bands, which are called active longitudes (ALs).
Active longitudes have been observed in several studies using different data bases \citep{temmeretal2006, chenetal2011, li2011, murakozyandludmany2012}.

Typically the ALs can be seen clearly only during a few solar rotations.
The longitudinal distribution of active regions becomes rather symmetric after a moderately long time interval, typically a few years.
 Accordingly, a simple way of calculating the longitudinal distribution of solar activity using the Carrington longitude of the active regions does not yield a persistently inhomogeneous distribution.

A few recent studies \citep{berdyuginaandusoskin2003, usoskinetal2005, zhangetal2011a, zhangetal2011b} using a dynamic, differentially rotating coordinate system have been able to show convincingly that the ALs are persistent structures that sustain for several tens of years, and possibly even longer.
A large fraction of various forms of solar activity are produced by regions which are themselves differentially rotating.
The persistence of ALs is clearly demonstrated in this reference frame, and the level of longitudinal asymmetry increases significantly higher than in any rigidly rotating system.
Moreover, the stronger forms of solar activity are more asymmetrically distributed than the weaker, more diffusive forms \citep{zhangetal2011a}.

Solar surface rotation has been studied for a long time using various forms of solar activity  \citep{balthasarandwohl1980, pulkkinenandtuominen1998, wangetal1988, brajsaetal2000}.
The secular deceleration of solar rotation was suggested by \citet{brajsaetal2006} and \citet{lietal2014}, while a secular acceleration trend was found by \citet{heristchiandmouradian2009}.
A north-south (N-S) asymmetry in solar rotation has been reported by many authors \citep{brajsaetal2000, javaraiahandkomm1999}.
Very recently, by studying the rotation of ALs of sunspots in the last twelve solar cycles it was found that the long-term evolution of the solar surface rotation  has a quasi-periodicity about 80-90 years \citep{zhangetal2013}.
Rotation was also found to be N-S asymmetric during most of this time period.
The consistent N-S asymmetry of solar surface rotation was confirmed by \citet{suzuki2014}.


The level of non-axisymmetry was found to increase when using shorter fit lengths  \citep{zhangetal2013}.
The average non-axisymmetry of sunspots increased from 0.280 for the 5-year running method to 0.464 for the 1-year method.
Since the rotation rates vary at a time scale of one year, the rotation parameters obtained for the shorter intervals represent the momentary rotation rate values of ALs more closely.
Accordingly, the ALs are more accurately determined from shorter intervals.

However, very short interval fits of a few solar rotations only may lose the continuous evolution of the rotational phase and thereby cause increased uncertainty when determining the rotation parameters.
The three-year fit interval length has been found to be the optimum length when both  satisfactory representation and continuous evolution are met \citep{zhangetal2013}.


Here we study the variation of solar rotation in recent years by studying the rotation parameters of ALs  of solar X-ray flares in 1977-2012.
To compare the results of flares with those of sunspots, we also study the rotation parameters determined by sunspots for the common period.
We present the yearly rotation rates at the average latitude of flares and of sunspots, respectively.

\section{Data and analysis method}

We study the solar X-ray flares of class-B and higher  observed by the NOAA GOES satellites during the period of 1977-2012.
Most X-ray flares were identified by simultaneous optical flare observations in the data prior to 2006.
However, a large fraction of the X-ray flares are without an accompanying optical flare since 2007.
No X-ray flare is observed together with an optical flare in 2011.
Fortunately, the flare location can be identified during the whole time period using the NOAA/USAF sunspot region number.
To keep the treatment consistent throughout the entire study period, we use the location of the sunspot region where the flare occurred as the location of the flare.
GOES X-ray flare data (ftp://ftp.ngdc.noaa.gov/STP/space-weather/solar-data/solar-features/solar-flares/x-rays/goes/) provides the NOAA/USAF sunspot group number where the flare is observed, but no location information for the sunspot groups.
The location information of sunspot groups can be retrieved from NOAA/USAF sunspot data (http://solarscience.msfc.nasa.gov/greenwch.shtml).
For sunspots we study the NOAA/USAF sunspot groups for the common period of solar flares.

The analysis method can be found in earlier studies \citep{usoskinetal2005, zhangetal2011a, zhangetal2013}, and is briefly described here.
Differential rotation of the solar surface is usually described as

\begin{equation}
      \Omega_{\phi} =\Omega_0 - {B}\sin^{2}{\phi},
\label{omega}
\end{equation}
where  $\Omega_{\phi}$ stands for the sidereal (all rotation rates are taken here to be sidereal) angular velocity at latitude $\phi$, $\Omega_0$ (deg/day) denotes the equatorial angular velocity, and $B$ (deg/day) describes the differential rotation rate.
The active longitudes are also assumed to follow the same form of differential rotation at their own specific values for the parameters $\Omega_0$ and $B$.

Assuming that the two ALs are at Carrington longitude $\Lambda_{0}$ and $\Lambda_{0} \pm 180^{\circ}$ and follow the differential rotation of solar surface, one can measure the distance between the longitude of a flare or sunspot group and the nearest AL.
The merit function can be defined as the mean square of these distances either without any weighting on flares or sunspots \citep{zhangetal2011a} or by weighting the flares with their normalized peak intensity and the sunspots with their normalized area. 
 The merit function $ \epsilon (\Lambda_{01}, \Omega_0, B)$ depends on the three parameters $\Lambda_{0}$, $\Omega_0$, and $B$ of the active longitudes.
 We use the least-square method to search for the best-fit parameters.
The 3-year running fit interval is used here, yielding the rotation parameters for the middle year.


We studied both the weighting method and no-weight method.
The results obtained with the two methods are very similar.
Thus, only the results of no-weight method are presented in this paper.

\section{Results}
\label{results}

\subsection{N-S asymmetry and recent speed-up of solar rotation }
\label{nsaymmetry}
\begin{figure*}
\centering
            \includegraphics[width=8.0cm, height=10.0cm]{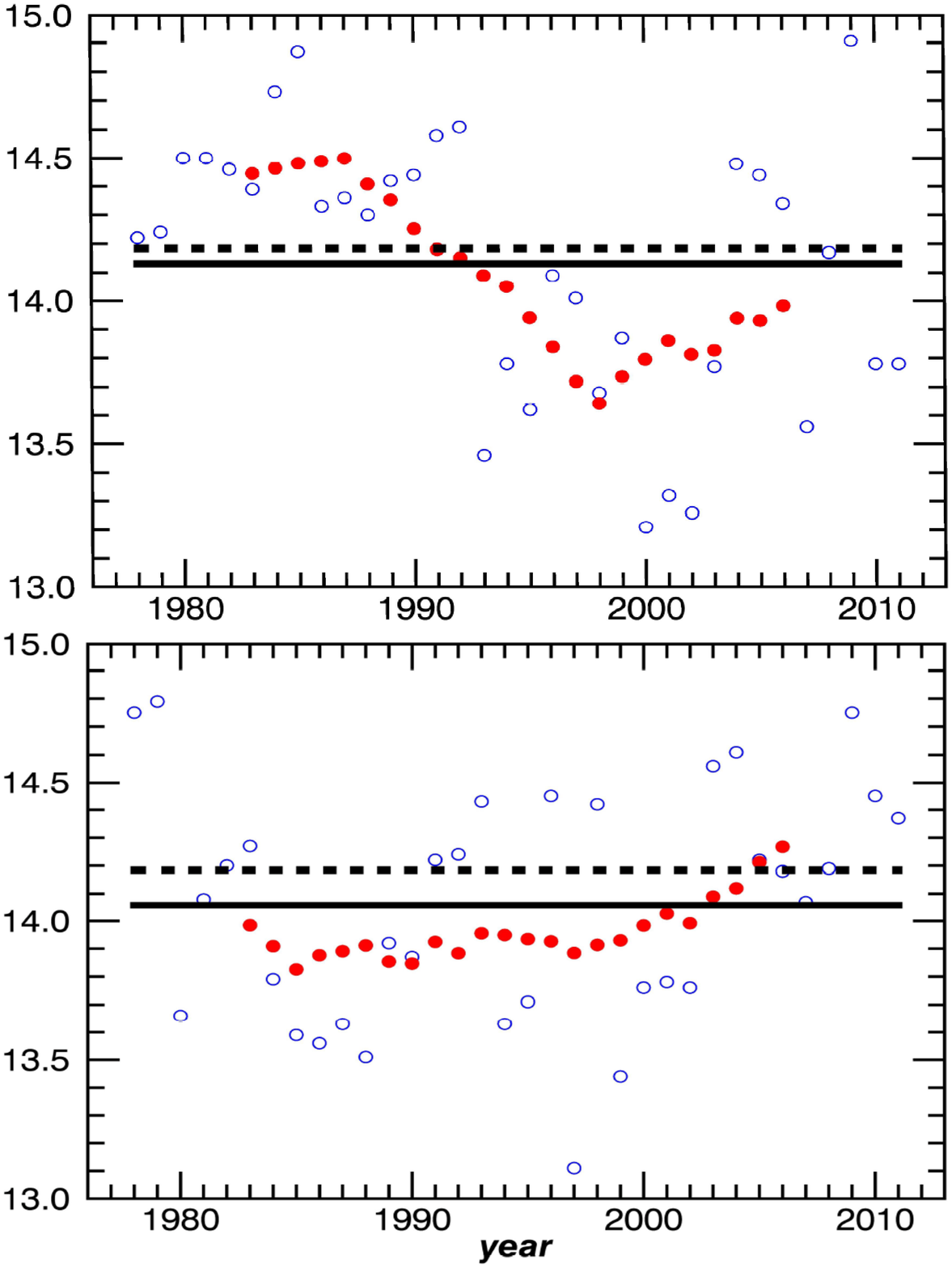}
          \includegraphics[width=8.0cm, height=10.0cm]{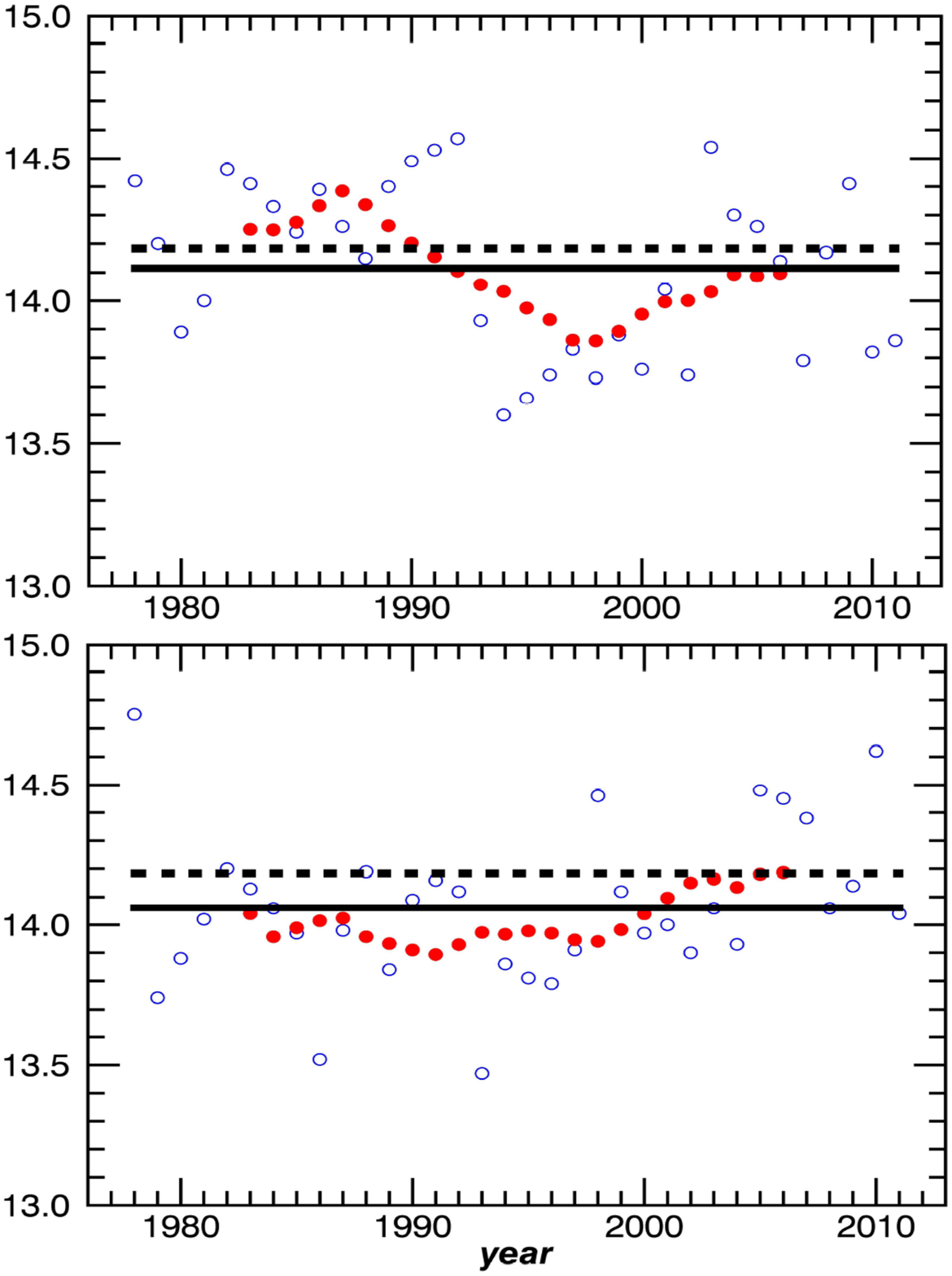}
\caption{Left-hand panels: Yearly values of $\Omega_{17}$ in the north (top) and south (bottom) for X-ray flares.
Open circles stand for the yearly best-fit values and red filled circles demonstrate the 11-point running mean values.
Solid horizontal line denotes the average  $\Omega_{17}$ over the entire study period, and dashed line the sidereal Carrington rotation rate.
 Right-hand panels: the same as in the left-hand but for sunspots.}
  \label{Figomega17}
\end{figure*}

We have calculated the yearly rotation rates $\Omega_{17}$ of ALs defined by formula (1)  at the reference latitude 17$^{\circ}$ for X-ray flares and sunspots, separately.
The values of $\Omega_{17}$ are presented in  Fig. \ref{Figomega17} (left-hand panels  for X-ray flares and right-hand panels for sunspots;
top panels for northern hemisphere and bottom panels for southern hemisphere).
Open circles depict the yearly values of $\Omega_{17}$ for the central year of each 3-year fit interval.
The 1$\sigma$ error for each $\Omega_{17}$ value is within $\pm$0.015 deg/day which is too small to be shown in the figure.
 To demonstrate the long-term variation pattern more clearly, the 11-point running mean (average over one solar cycle) is shown in the figure  as red filled circles.

\begin{figure}
\centering
      \includegraphics[width=8.0cm, height=5.0cm]{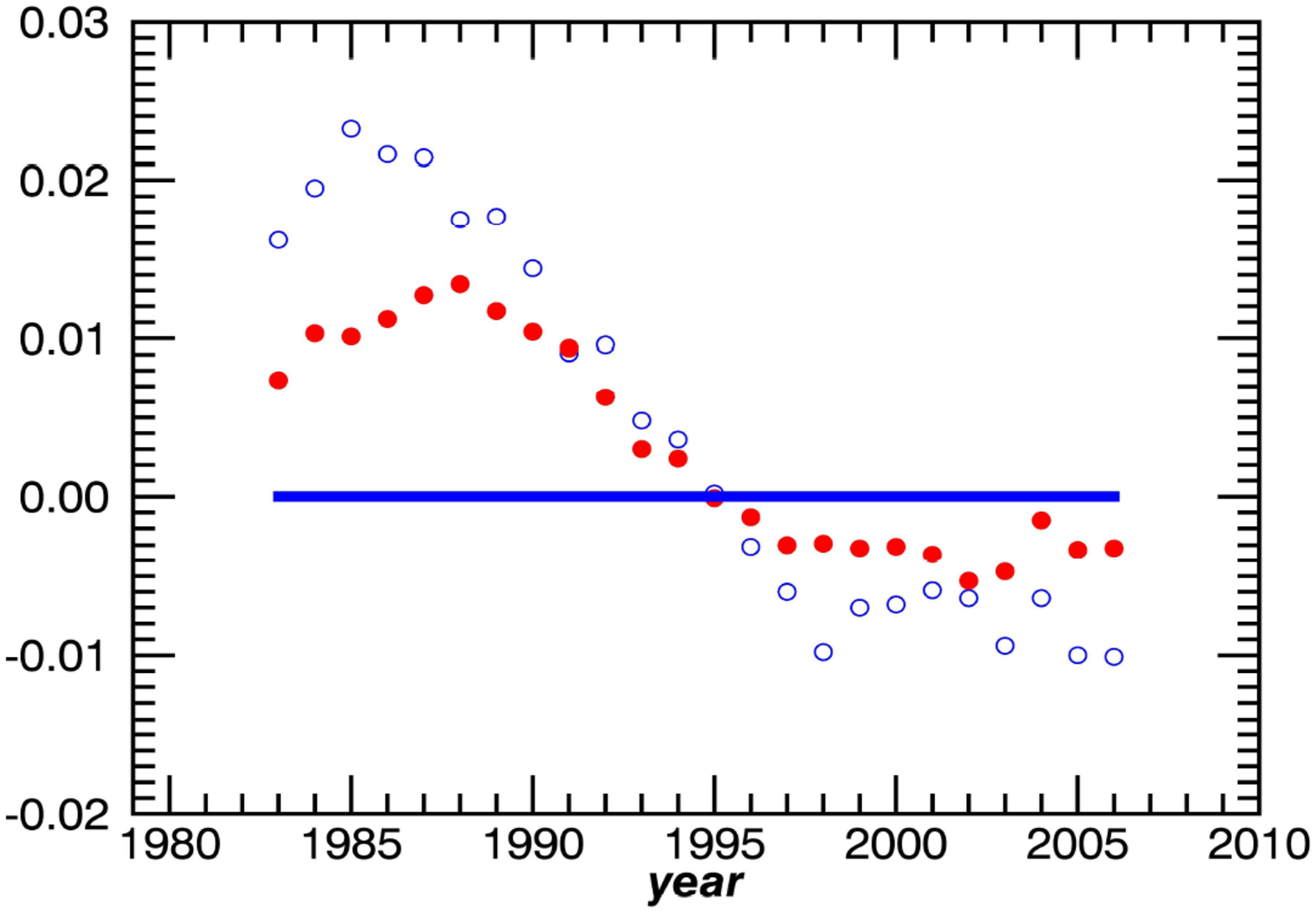}
\caption{Yearly values of N-S asymmetry of solar rotation at latitude 17$^{\circ}$ obtained for X-ray flares (open circles) and sunspots (filled circles).}
  \label{FigNSasymmetry}
\end{figure}

The rotation rates of $\Omega_{17}$ in the two hemispheres obtained with X-ray flares depict little correlation from 1977 until pre-2000.
In the 1980s the north rotates faster than the south but slows down in the 1990s even below the southern rate.
During this time the southern rotation rate remains rather constant.
The rotation rate has been speeding up in both hemispheres since pre-2000, with the southern hemisphere rotating faster than the north.
The rotation evolution obtained for sunspots depicts a notably similar pattern as flares.
However, the recent rotation of the northern hemisphere according to sunspots is somewhat faster than that according to flares.
Therefore, the hemispheric asymmetry in the rotation rate is smaller for sunspots.
This is probably caused by a few large and flare productive active regions which rotate rather slowly in the northern hemisphere.
Figure \ref{FigNSasymmetry} depicts the N-S asymmetry ((N - S)/(N + S)) of solar rotation at the latitude of 17$^{\circ}$ obtained for X-ray flares (open circles) and sunspots (filled circles).
The evolution of the hemispheric asymmetry of solar rotation is quite similarly depicted by flares and sunspots:
while the asymmetry has been fairly constant and negative since the late 1990s, it was strongly positive in the 1980s and decreased rapidly from mid-1980s to mid-1990s.


\begin{figure}
\centering
       \includegraphics[width=8.0cm]{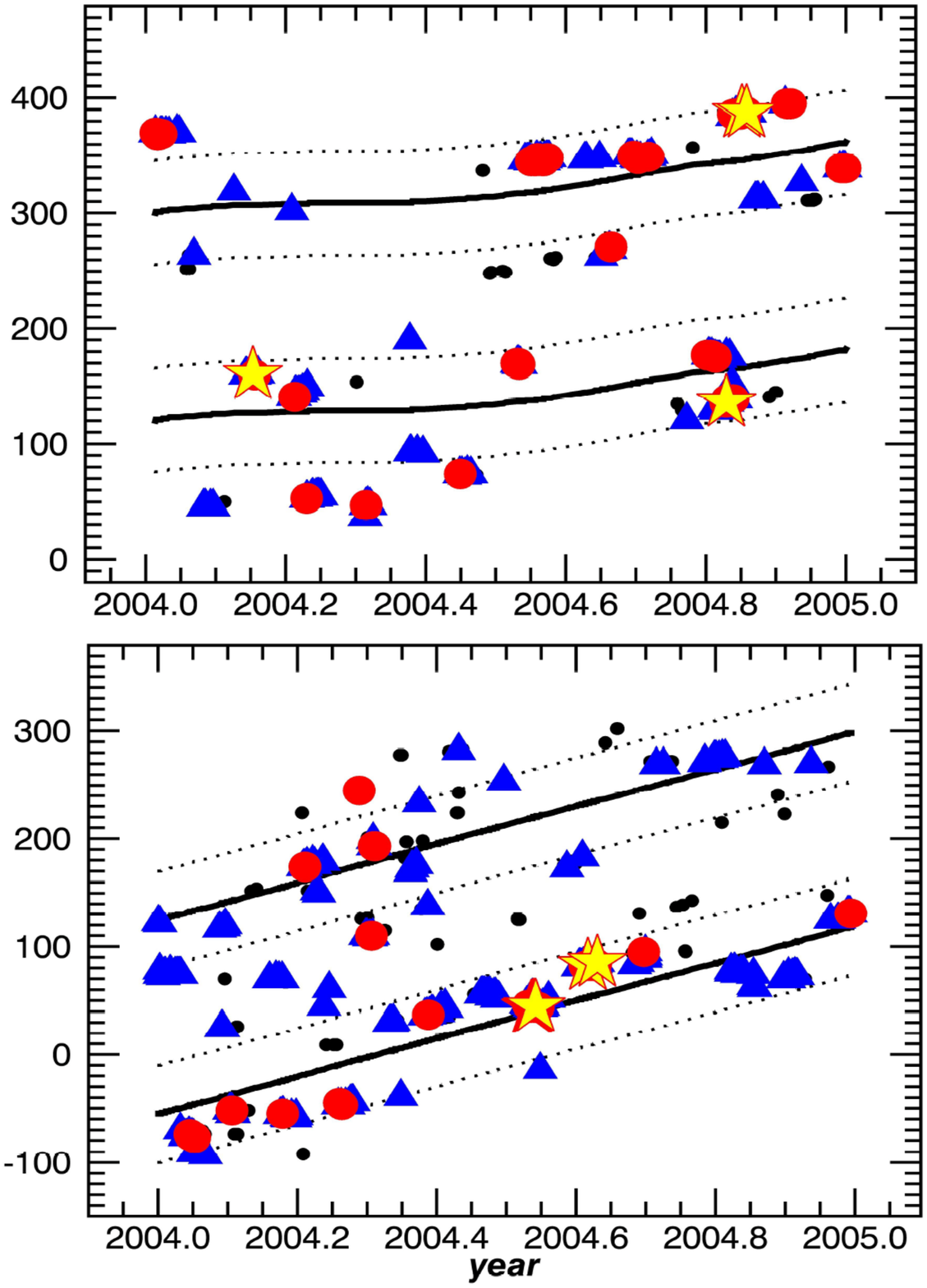}
  \caption{Migration of ALs  in northern (top) and southern hemispheres (bottom) in 2004. Black dots stand for B-flares, triangles C-flares, red dots M-flares and stars X-flares.
  The two solid lines depict the migration of the two ALs from the beginning to the end of 2004 with 45$^{\circ}$ extensions on each side denoted by dotted lines. }
  \label{FigAL2004}
\end{figure}

 The fact that the southern hemisphere rotates faster than the north can also be seen in the migration of ALs in the two hemispheres.
Figure \ref{FigAL2004} demonstrates the migration of ALs in the Carrington reference frame in the northern (top) and southern (bottom) hemispheres in 2004.
Black dots denote the B-flares, triangles C-flares, red dots M-flares and stars X-flares.
The two solid lines in each panel depict the migration of the two ALs from the beginning to the end of 2004.
The dotted lines on either side of the two solid AL lines denote the 90$^{\circ}$ ($\pm$45$^{\circ}$) regions around the two ALs.

One can see that in the northern hemisphere the longitudinal location of one AL increases gradually from 120$^{\circ}$ to 180$^{\circ}$-190$^{\circ}$ in the Carrington reference frame in 2004 yielding the total increase of 60-70$^{\circ}$ in one year, while in the southern hemisphere the total increase in this year is about 160$^{\circ}$-170$^{\circ}$.
This indicates that both hemispheres rotate  consistently faster than the Carrington reference frame (as shown in Fig. \ref{Figomega17}) and the southern hemisphere rotates significantly faster than the north (as depicted by the negative values in Fig. \ref{FigNSasymmetry}).

\subsection{Non-axisymmetry of ALs}
\label{nonaxisymmetry}

We define the measure of non-axisymmetry $\Gamma$ as follows
\begin{equation}
           \Gamma=\frac{N_{1}-N_{2}}{N_{1}+N_{2}},
   \end{equation}
where $N_{1}$ and $N_{2}$ denote the number of solar flares that appeared within ($N_{1}$) or outside ($N_{2}$) the two
AL regions, which are taken here as the two 90$^{\circ}$-longitude bands, depicted in Fig.  \ref{FigAL2004}.

The averaged yearly values of $\Gamma$ for sunspots and for each flare class are listed in Table 1 (online).
The asymmetries for the different flare classes were calculated using the same rotation parameters of the (common) ALs.
The corresponding fractions of flares in the ALs are also listed in Table 1.
The non-axisymmetry of flares in the southern hemisphere increases systematically with flare class from 0.422 (71.1\%)  of B-flares to 0.679  (83.9\%)  of X-flares.
In the northern hemisphere, the increase of non-axisymmetry with flare class is less dramatic and less systematic.
The non-axisymmetry 0.488   (74.4\%) of X-flares is slightly lower than the non-axisymmetry  0.506    (75.3\%) of M-flares.
This indicates that the rotation rates of these active regions in the northern hemisphere that produce X-class flares and those that produce other flares of other classes have somewhat larger differences than in the southern hemisphere.
All the non-axisymmetry values of flares in both hemispheres are larger than the non-axisymmetry of sunspot.
This is in agreement with the earlier finding that more powerful solar activity attends to appear closer to the two ALs \citep{zhangetal2011a, zhangetal2011b}.

\section{Discussion and conclusion}
\label{discussion}

The relationship between solar rotation and sunspot activity is a foundamental problem.
It is known that sunspot activity presents a secular increase during the 20th century leading to the so-called Modern Maximum (MM) \citep{solankietal2004}.
\citet{brajsaetal2006} and \citet{kitchatinovetal1999} found a secular deceleration of solar rotation, implying a negative correlation between solar rotation rate and sunspot activity.
However, a secular acceleration in solar rotation has also been proposed \citep{heristchiandmouradian2009, lietal2014}.
Moreover, \citet{ribes1993} claimed that the solar surface rotation at 20$^{\circ}$ latitude was 6\% slower during Maunder Minimum than in modern times.
These latter studies suggest that the correlation between solar rotation rate and sunspot activity is positive.

 We find that the northern and southern hemispheres both have been speeding up since late 1990s - the ending phase of MM.
 The activity level in cycle 23 significantly reduced compared with other cycles during the space era.
The recent minimum between cycles 23-24  lasted exceptionally long, and various solar activity measurements reached abnormally low values.
Solar wind density and the heliospheric magnetic field intensity reduced by nearly 1/3, both reaching uniquely low levels since the measured time of about 50 years \citep{cliverandling2011, jianetal2011}.
The sunspot activity in cycle 24 is even more dramatically reduced and matches the low level in the beginning of the 20th century.
This supports the negative correlation between solar rotation rate and sunspot activity.

The recent speed-up of solar rotation was also found in our earlier study of sunspots. 
\citet{zhangetal2013} studied the long-term evolution of solar rotation by analyzing the ALs of sunspots since 1870s.
Besides the recent speed-up period starting at pre-2000, there were several periods where one of the two hemispheres was accelerating, but very few when both hemispheres were speeding up.
Only two such periods were found, one during cycle 12 in 1880s and one during cycle 14 in the early 1900s (see Fig. 1 in \citet{zhangetal2013}), but both were shorter than the period of recent speed-up.
Note that cycles 14 and 12 were the two lowest cycles at the turn of the centuries 100 years ago and that the minimum between cycles 14-15 lasted exceptionally long, similarly to the previous cycle.
During the recent minimum, 817 days were observed without sunspots.
 During the minimum between  cycles 14-15 more than one thousand days were recorded as spotless \citep{cletteandlefevre2012}.

The recent speed-up of solar rotation coincided with the breakdown of the mutual relationships among several solar activity and geo-activity indices. 
A large divergence  was observed at about 2001-2002 between the sunspot numbers and several UV/EUV flux proxies, including the F10.7 radio flux  \citep{floydetal2005, lefevreandclette2011, lukianovaandmursula2011, kane2003, leanetal2011, liuetal2011}.
These changes can be understood in terms of the changes in sunspot distribution and the recent vanishing of small sunspots  \citep{cletteandlefevre2012}.

To summarize, we find that both solar hemispheres have increased their rotation rate since the late 1990s until recent years.
Also, the rate of increase is fairly similar in the two hemispheres so that the hemispheric asymmetry in rotation rates has been roughly constant, with southern rotation being slightly faster.
This period of recent solar speed-up coincides with the decline of the Modern Maximum (period of exceptionally high activity during most of the 20th century), which is evidenced by the overall reduction of sunspot activity, vanishing of small sunspots, decreasing solar wind density and magnetic field etc.
We also note that similar, although shorter and less significant periods of both hemispheres speeding up were only found during the two lowest solar cycles 12 and 14 at the turn of the 19th and 20th century.
These results strongly suggest that there is, at least momentarily, a negative correlation between solar surface rotation rate and sunspot activity.
However, the causes to these results and their possible implications to, e.g., solar dynamo, remain for subsequent studies to be solved.


\begin{acknowledgements}
The research leading to these results has received funding from the European Commission's Seventh Framework Programme (FP7/2007-2013) under the grant agreement eHeroes (project n$^{\circ}$ 284461, www.eheroes.eu).
We also acknowledge the financial support by the Academy of Finland to the ReSoLVE Centre of Excellence (project no. 272157).

\end{acknowledgements}



\begin{table}
\begin{center} 
\caption[]{Average non-axisymmetries of sunspots and the different flare classes obtained with the best-fit parameters of ALs.
Percentage of flares and sunspots within the two ALs, $N_1/(N_1+N_2)$, are given in parenthesis.}
\begin{tabular}{ccccc}
 \hline
     \noalign{\smallskip}
   &  north   & south  \\
 \hline
  \noalign{\smallskip}
   &  $\Gamma$  &  $\Gamma$   \\
     \noalign{\smallskip}
      \hline
          \noalign{\smallskip}
sunspots   & 0.313  (65.7\%)   & 0.330 (66.5\%)     \\
B-flares    &  0.422  (71.1\%)   & 0.422 (71.1\%)     \\
C-flares    & 0.474   (73.7\%) & 0.425 (71.2\%)         \\
M-flares   & 0.506    (75.3\%) & 0.572 (78.6\%)      \\
X-flares    & 0.488   (74.4\%)  & 0.679 (83.9\%)        \\
  \hline
  \end{tabular}
\end{center}
\label{non-axisymmetry}
\end{table}


\end{document}